# Charge-Transfer State Dynamics Following Hole and Electron Transfer in Organic Photovoltaic Devices

*Artem A. Bakulin,[1,2]\* Stoichko D. Dimitrov,[3] Akshay Rao,[2] Philip C. Y. Chow,[2] Christian B. Nielsen,[3] Bob C. Schroeder,[3] Iain McCulloch,[3] Huib J. Bakker,[1] James R. Durrant,[3] and Richard H. Friend [2]\**

[1] FOM institute AMOLF, Science Park 104, Amsterdam, The Netherlands

[2] Cavendish Laboratory, University of Cambridge, JJ Thomson Ave, Cambridge CB3 0HE, United Kingdom

[3] Centre for Plastic Electronics, Department of Chemistry, Imperial College London, Exhibition Road, London SW7 2AZ, United Kingdom

Corresponding Authors: * a.bakulin@amolf.nl; rhf10@cam.ac.uk ;

**Abstract**

The formation of bound electron-hole pairs, also called charge-transfer (CT) states, in organic-based photovoltaic devices is one of the dominant loss mechanisms hindering performance. While CT state dynamics following electron transfer from donor to acceptor have been widely studied, there is not much known about the dynamics of bound CT states produced by hole transfer from the acceptor to the donor. In this letter, we compare the dynamics of CT states formed in the different charge-transfer pathways in a range of model systems. We show that the nature and dynamics of the generated CT states are similar in the case of electron and hole transfer. However the yield of bound and free charges is observed to be strongly dependent on the $HOMO_D$-$HOMO_A$ and $LUMO_D$-$LUMO_A$ energy differences of the material system. We propose a qualitative model in which the effects of static disorder and sampling of states during the relaxation determine the probability of accessing CT states favourable for charge separation.

MAIN TEXT

An organic-based photovoltaic (OPV) cell [1] consists of a nanostructured blend of two materials, an electron donor (D) and an electron acceptor (A), sandwiched between oxide [2] or metal electrodes. Upon illumination, the incident photons are absorbed by one of the materials and converted to an intra-molecular excitonic state that can subsequently dissociate into a pair of spatially separated charges at the D-A interface.[3] When the excitonic state is localised on the donor material, the charge separation involves charge exchange between the lowest unoccupied molecular orbitals (LUMO) of D and A. This process, usually referred as electron transfer, is driven by the $LUMO_D$-$LUMO_A$ energy difference. When the exciton is localised on the acceptor, the charge separation involves highest occupied molecular orbitals (HOMO).[4,5] Such process, usually called hole transfer, is driven by the energy difference between the HOMO of the A and D material and depends on the coupling between their valence bands. In either case, charge transfer leads to an intermolecular charge-transfer (CT) state consisting of an electron and a hole that are located on the A and D respectively.[6,7] In the CT state, the charges still interact with each other as a result of the Coulomb forces acting across the D-A interface. The properties and dynamics of the CT states are known to be critical for the photoconversion efficiency, as bound charges are more likely to recombine and thus may not contribute to the device photocurrent.[8] There are a large number of studies of the CT state dynamics reported for organic materials and devices.[9-23] [24] While most of these studies have been focused on the CT states that are formed after electron transfer, little attention has been given to the properties of the CT states that result from frequently disregarded [24] hole transfer and to the role of $HOMO_A$-$HOMO_D$ driving energy for charge separation.[25,26]

In this letter we use a novel IR pump-push photocurrent (PPP) spectroscopy technique [27,28] to compare the dynamics of CT states produced via hole or electron transfer. We study a set of material systems with different band offsets and driving energies for charge separation and observe that the dynamics of the hole-transfer generated CT states are very similar to the dynamics of the electron-transfer generated CT states indicating the similarity in the nature of the states. However, the yields of bound and free charges strongly depend on the $HOMO_D$-$HOMO_A$ or $LUMO_D$-$LUMO_A$ energy differences of the material system.

The systems under study were three different polymer-fullerene blends. **Figure 1a** shows the chemical structures of the donor and acceptor molecules involved, and **figure 1b** presents the previously reported positions of their energy levels.[29-31] Although band diagrams do not provide comprehensive information about charge-separation pathways and CT-state position, they emphasise the difference in the energy provided for charge separation in the studied systems. We used well-studied donor-acceptor blends with similarly large band offsets both for the hole and the electron transfer (MDMO-PPV:$PC_{70}BM$ and PCPDTBT:$PC_{70}BM$) and a blend of BTT-DPP with $PC_{70}BM$, where the $HOMO_D$-$HOMO_A$ energy difference is much larger than the $LUMO_D$-$LUMO_A$ offset.[31,32] We note that the energy values of the polymers and $PC_{70}BM$ are obtained from different sources and posses an uncertainty, which we estimate to be ~0.1 eV for the polymers and even higher for the $PC_{70}BM$. Those values are shown here to illustrate the general trend in the variations of the driving energy for charge separation in the material systems used.

**Figure 2a,b,c** presents the absorption spectra of donor and acceptor for each blend together with the excitation wavelengths used in the PPP experiments. Note that we have used $PC_{70}BM$ because it shows substantial optical absorption, particularly near 520 nm, in contrast to $PC_{60}BM$. The spectra were measured for films of each material separately and later normalised to obtain a reasonable estimate of the donor and acceptor contributions to the absorption in the blend at different wavelengths. For all material systems the absorption bands of D and A are shifted with respect to each other, which implies that the D and A molecules can be selectively excited with relatively high D vs. A contrast using light pulses centred at ~520 nm or ~700 nm. We used this method to selectively initiate either electron or hole transfer [26] and study the dynamics of the resulting CT states in the photovoltaic cells. For MDMO-

PPV:PC$_{70}$BM we excite the polymer with 550 nm light after which the CT generation proceeds through the electron-transfer pathway. For the same system, excitation of the fullerene acceptor with 680 nm light produces CT states via hole transfer. The PCPDTBT and BTT-DPP polymers both absorb at longer wavelength than PC$_{70}$BM and therefore we use 700 nm pump light to excite the polymer donor (electron transfer) and 500-520 nm light to excite the fullerene acceptor (hole transfer). In our experiments we presume that energy transfers from donor to acceptor and vice versa do not play a significant role in photophysics, because electron and hole transfers happen at the ultrafast timescale (30-45 fs),[24] allowing them to compete efficiently with energy transfer in the material.

We studied the dynamics of charge generation with pump-push photocurrent spectroscopy,[27,33] a technique that is ideally suited to probe the presence of strongly bound CT states in an operational photovoltaic device. In PPP experiments (**figure 3a**), a working OPV cell is exposed to a visible (~520 nm or ~700 nm) pump pulse leading to bound and free charge carriers via hole or electron transfer process. The generated free carriers create a 'reference' photocurrent output $J$ of the device. After a delay, the cell is illuminated by an IR (2200 nm) push pulse which is absorbed selectively by the charged polaronic states on the polymer chains,[34] providing them with an extra energy of ~0.5 eV. If the charges are free, their dynamics are hardly influenced by the excess energy as free charges quickly thermalise (~100fs),[27] thus rapidly returning to the 'free carrier' state similar to the one they were in before the excitation. In contrast, for the charges that are bound in interfacial CT states, the excess energy can have a significant effect on the dynamics, as the excess energy can lead to dissociation of these CT states. The dissociation gives an additional photocurrent $\delta J$. The normalized change in current $\delta J/J$ thus forms a measure of the relative amount of bound CT states in the cell. By measuring $\delta J/J$ as a function of the delay between the push and the excitation pulses we get information on the dynamics of the bound CT states. These dynamics are fingerprints of the CT state electronic structure and are used here to identify the variation in the nature of interfacial charge carriers.

**Figures 3b,c** compare the PPP kinetics for MDMO-PPV:PC$_{70}$BM and PCPDTBT:PC$_{70}$BM cells after excitation at 500--550 nm and 680-700 nm. In all measurements, when the push pulse arrives before the pump, the effect on the photocurrent is negligible, because there are very few charges in the cell to be influenced by the push pulse. When the push pulse arrives after the pump, the $\delta J/J$ value manifestly increases and then gradually decays. The fact that the absolute values of $\delta J/J$ are different for the presented molecular systems can be related to the different yield of bound charges as well as to the different absorption cross-sections of CT states at the push-photon energy. The observed increase and decay reflect the fast (~1 ps) generation and ~300 ps recombination of bound CT states in the photovoltaic cell, in good agreement with the previously reported photoluminescence data.[35] For each sample, the PPP kinetics are observed to be very similar for the different excitation wavelengths with only the amplitudes of the responses being slightly different. The similarity between the kinetics indicates that the bound CT states formed in both cases are almost identical and, therefore, do not depend on the mechanism (electron or hole transfer) by which they are formed. Note, that although the contrast between donor and acceptor excitation for 500-550 nm pump is not very high, at longer pump wavelength only one component of the blend is excited. This allowed us to draw a qualitative conclusion about the similar charge dynamics in both charge-generation pathways.

The difference in amplitudes of PPP transients observed at different excitation frequencies indicates that the charge transfer pathway does determine the probability to form bound CT states. The fundamental mechanism behind these probability variations may involve different couplings between D, A and CT state electronic structures, dissimilar morphology and dielectric environment of charge separation sites, or the HOMO$_A$-HOMO$_D$/LUMO$_A$-LUMO$_D$ energy difference that drives the formation of the CT states. For MDMO-PPV:PC$_{70}$BM the driving energy is higher for electron transfer than for hole transfer, and the electron-transfer related PPP transient displays a lower amplitude. For PCPDTBT:PC$_{70}$BM the driving energy for hole transfer exceeds that for electron transfer and, consistently, the amount of bound CT states generated after hole transfer is lower. We thus find that the yield of bound CT states decreases when the energy difference driving the charge transfer increases. This observation may be explained from the fact that a larger energy difference will result in a larger excess energy of the transferred charge and thus a lower probability for formation of bound CT states. Alternative explanation can be that sample areas with different local morphologies are accessed by a pump of different colour. The effect of energy-level offsets on charge separation has been investigated before and it was found to show a consistent dependence on the energy offset for a wide range of molecular systems.[11,36] To investigate the effect of the driving energy in more detail we performed ultrafast measurements on BTT-DPP:PC$_{70}$BM devices for which the difference in the driving energy for electron- and hole-transfer scenarios is dramatically different.[31]

**Figure 4** presents the results of PPP experiment on a BTT-DPP:PC$_{70}$BM cell. Similarly to the previously discussed systems we find the kinetics of generation and recombination for the bound CT states to be similar for excitations in the regions of the polymer and the fullerene. However, the yield of bound CT states is much higher in the case of polymer (donor) excitation at 700 nm. This finding can be explained from the very low LUMO$_D$-LUMO$_A$ energy difference in this material which provides a sufficient driving force for electron transfer but not for long-range charge separation.[31,32]

The origin of the low-amplitude PPP signal observed in BTT-DPP:PC$_{70}$BM device after excitation at 520 nm requires additional investigation. The contrast between the acceptor and donor excitations at 520 nm is not very high (the absorption cross-section of the acceptor is only ~3 times higher than that of the donor) and, taking into account the observed much higher yield of bound charges after donor excitation, both electron and hole transfer processes may contribute similarly to the generation of the bound CT states. To distinguish between the contributions to the charge generation from different charge transfer scenarios we performed novel PPP anisotropy experiments on this system. Transient anisotropy measurements have been used before to distinguish between electron and hole transfer contributions to the charge generation in MDMO-PPV:PCBM blends.[26,37] It was shown that electron-transfer process demonstrates a noticeable conservation of the polarisation memory due to the correlation between the excitonic, CT and polaronic transition dipoles of the polymer.[38-41] In contrast, the polarization of the fullerene excitation has little correlation with the polarization of the polymer's polaronic transition, and thus the anisotropy following hole transfer will be negligibly low.

The **inset** in **figure 4** shows the PPP anisotropy for BTT-DPP:PC$_{70}$BM measured at different excitation wavelengths. The anisotropy is constant during the first 20 ps after excitation demonstrating that in both cases the generated CT states are immobile and do not relocate between different polymer chains within their lifetime. The different levels of anisotropy we associate with the different charge generation pathways. With the pump set at 700 nm, mostly the polymer donor is excited and the corresponding anisotropy reaches a relatively high value of 0.15. After excitation at 520 nm, the anisotropy is approximately half that value. We expect that the anisotropy following photoexcitation of the fullerene will be very low.[26] We therefore consider that the contribution of hole transfer to the observed PPP response is around 50%. Taking into account the difference between the observed amplitudes of the isotropic transients, we estimate the yield of bound CT states to be about 10 times lower for hole transfer than for electron transfer in the BTT-DPP:PC$_{70}$BM blend. This difference follows from the much larger HOMO$_D$-HOMO$_A$ energy difference compared to the LUMO$_D$-LUMO$_A$ energy difference.

The PPP data presented above represent a direct probe of bound CT state dynamics in model polymer-fullerene systems and the effect of energy offsets on this process, an area which has been intensely debated over the past few years. For instance, it has been shown that for a range of systems the yield of charges in polymer/fullerene blends correlate with the energetic driving force for charge separation, which was interpreted in terms of competition between the CT-relaxation and charge-separation kinetics.[11] A high excess energy leads to the population of intermediate "hot" CT states with enough thermal or electronic excess energy to facilitate charge separation, while a low excess energy leads to bound CT state formation. In contrast, a range of studies have demonstrated that excess energy is not the only factor determining charge separation, as the formation of free charges was also observed from the relatively low-lying CT states,[39,42-44] which, however, may still be high enough in energy to facilitate dissociation.[45] It has also been observed that the relaxation of the CT states occurs on ultrafast timescales (<200fs) and thus charge separation cannot be explained as a classical kinetic process.[27] These and other observations have demonstrated the importance of electron-hole pair delocalisation or/and an increased acceptor electron mobility (often associated with increased wavefunction delocalisation) for the charge separation efficiency.[15,22,27,36]

Here we combine the energy offset and delocalisation requirements for charge separation and produce a consistent framework to explain charge separation across a range of systems. The results of this paper demonstrate that the yield of bound CT states increases with the decreasing amount of driving energy, the dependence being much stronger for the material system with a small band offset for charge separation. The probability for a particular CT state to dissociate into a pair of free charges is dictated by the properties of the material system (delocalisation, polaronic effects, local morphology, etc.) and does not depend on the way the CT state was populated (electron or hole transfer). However, the probability to populate CT states that are favourable for charge separation will depend on the driving energy as this energy determines the manifold of CT states energetically accessible following the charge transfer step. This is shown in **figure 5(a,b)**; in the case of material systems with 'low' (<0.2eV) band offsets (figure 5b), the static disorder dominates the system by determining a distribution of low-energy CT states that can be populated after the initial charge transfer step. Only a few of these low lying CT states, with the required level of delocalisation, are therefore favourable for the free-carrier formation. In contrast, when 'high' driving energies are involved (figure 5a), the static-distribution effects are overtaken by the large number of CT states that can be sampled during the thermalisation. The band offsets of the investigated material systems (and in majority of organic PV systems studied until now) are much larger than the thermal energy making the CT-state sampling excitation-wavelength and temperature insensitive; this is in agreement with recent reports that showed the generation of free carriers to be independent on temperature.[44,46]

In summary, we have compared the dynamics of CT states produced in hole-transfer and electron transfer charge-separation pathways in three organic PV cells. We observed that the kinetics of CT state recombination are very similar for both charge generation mechanisms, which indicates that the nature of the generated CT states does not depend on the charge-transfer pathway. However, we observed that the yield of bound and free charges strongly depends on the particular driving energy provided by $HOMO_D$-$HOMO_A$ (hole transfer) or $LUMO_D$-$LUMO_A$ (electron transfer) of the material system. The yield of bound CT states strongly increases with decreasing energy difference driving the charge transfer. The results fit well with a qualitative model of charge separation where the effects of static disorder and sampling of the CT manifold during the thermalisation influence the probability to access the CT states favourable for free carrier generation.

**METHODS**

Device preparation and characterisation: $PC_{70}BM$ was used as purchased from Nano-C. Solutions of MDMO-PPV: $PC_{70}BM$ (1:1), PCPDTBT: $PC_{70}BM$ (1:2) and BTT-DPP: $PC_{70}BM$ (1:3) were prepared in chloroform, chlorobenzene and o-dichlorobenzene respectively. ITO substrates with sheet resistance 15 $\Omega sq-1$ (PsioTec Ltd, UK) were sonicated in detergent (acetone, isopropanol) before treating in an oxygen plasma asher. PEDOT:PSS was then spin-coated over the ITO substrates at 3000 rpm and dried on a hot plate at 150°C in air for 20 min. MDMO-PPV: $PC_{70}BM$ and PCPDTBT: $PC_{70}BM$ active layers were spin-coated over the PEDOT:PSS layer in a nitrogen glove box, followed by deposition of Al electrodes (100nm). For BTT-DPP: $PC_{70}BM$ devices, the active layer was spin-coated in air and transferred to a nitrogen glove box, where LiF/Al (1 nm/100 nm) electrodes were thermally evaporated under vacuum. Final device active layer was 0.045 cm$^2$. All devices were encapsulated under nitrogen atmosphere and did not show any sign of degradation during the measurements.

Pump-Push Photocurrent Spectroscopy: A regenerative 1 kHz Ti:Sapphire amplifier system (Coherent, Legend Elite Duo) was used to pump both a broadband non-collinear optical amplifier (Clark) and a 3-stage home-built optical parametric amplifier (OPA) to generate visible pump pulses (±20 nm bandwidth) and infrared push pulses (2200±100 nm), respectively. ~1 nJ pump and ~0.5 µJ push pulses were focused onto a ~1 mm$^2$ spot on the device. The reference photocurrent from a photodiode was detected at a pump repetition frequency of 1 kHz by a lock-in amplifier. The push beam was mechanically chopped at ~380 Hz, and its effect on the photocurrent was detected by a lock-in amplifier. For anisotropy measurements the polarization of push beam was set by a wire-grid polarizer (1:100 extinction) and the polarization of visible pump was rotated using a combination of an achromatic half-wave plate and a thin film polariser (1:300 extinction). To avoid experimental artefacts, the intensity dependence of the signal was measured and checked for multi-photon contributions.

**ACKNOWLEDGEMENTS**

We thank the EPSRC for financial support. We also thank Nurlan Tokmoldin for assistance with device fabrication. A.A.B. acknowledges a VENI grant from the Netherlands Organization for Scientific Research (NWO). A.R thanks Corpus Christi College for a Research Fellowship.

**FIGURE LEGEND**

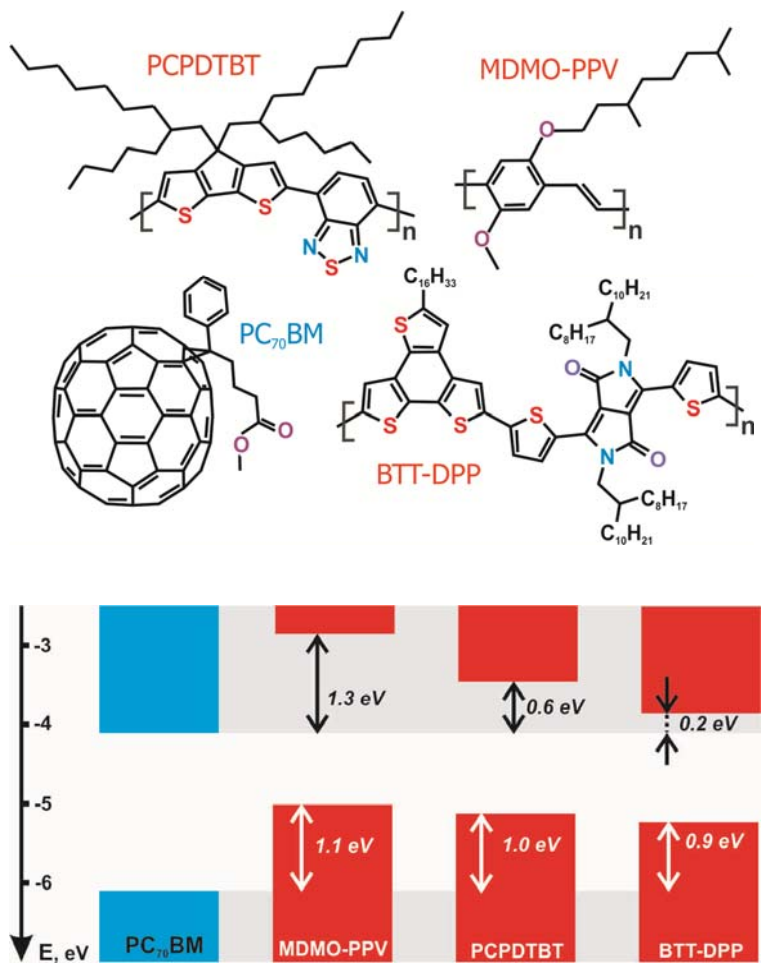

Figure 1. Chemical structures of the materials used, and estimates of the HOMO and LUMO energy levels. The arrows indicate the estimated driving energies for electron and hole transfer.

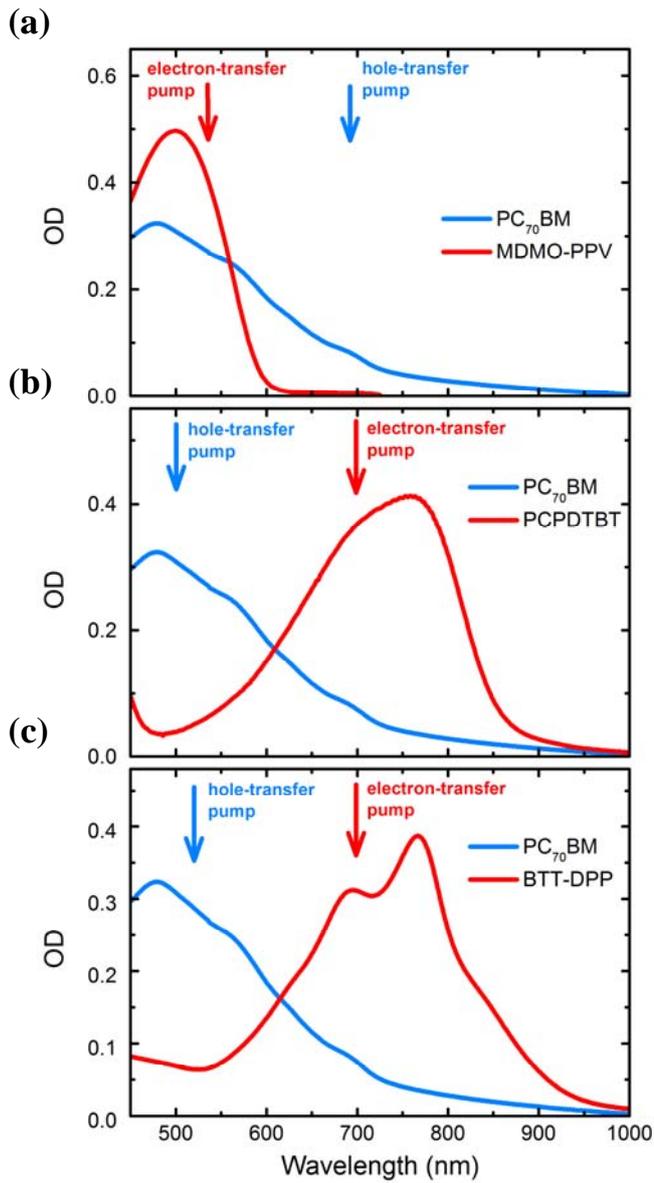

Figure 2. Absorption spectra of the materials used. The arrows show the frequencies of the pump pulses used in the pump-push photocurrent experiments.

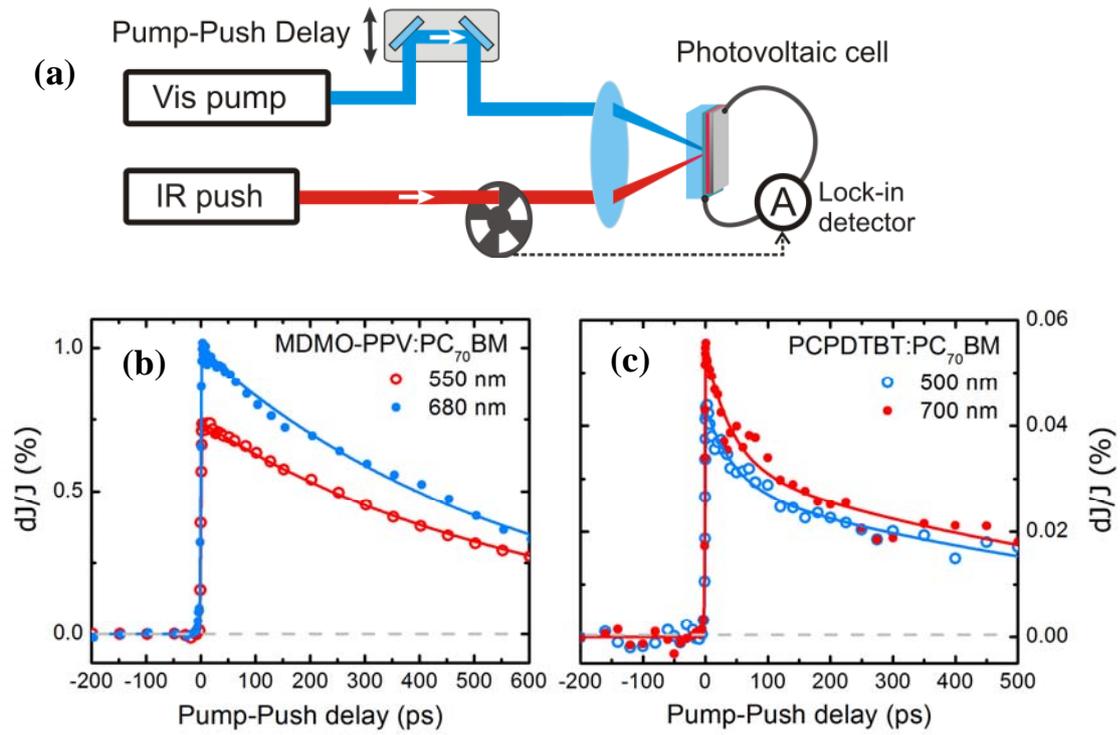

Figure 3. Top panel: schematic picture of the pump-push photocurrent spectroscopy setup. Lower panels: results of pump-push photocurrent ($\delta J/J$) measurements on MDMO-PPV:PC$_{70}$BM and PCPDTBT:PC$_{70}$BM devices at different excitation wavelengths leading to CT state generation through electron transfer or hole transfer. The lines are (bi-)exponential fits convoluted with the 150 fs instrument function.

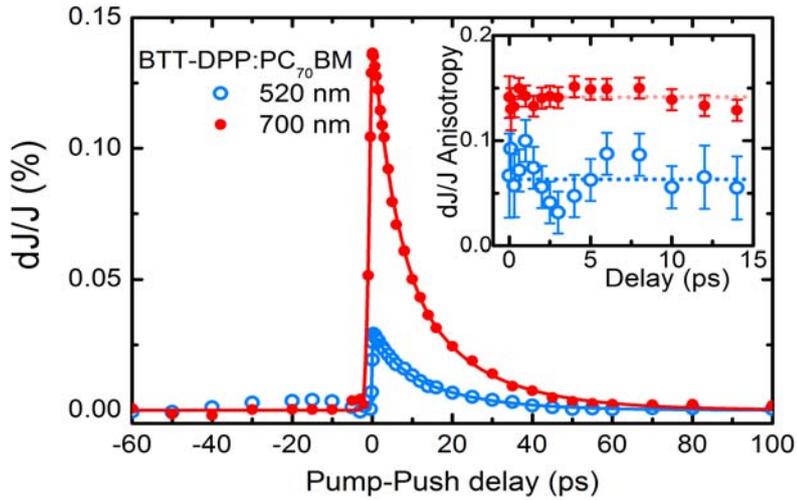

Figure 4. Isotropic component of the pump-push photocurrent response of a BTT-DPP:$PC_{70}BM$ device excited at 520 nm (mostly $PC_{70}BM$ excitation) and at 700 nm (mostly BTT-DPP excitation). The inset shows the corresponding anisotropy dynamics. For 700 nm the observed bound CT states originate from electron transfer, for 520 nm the bound CT states result both from hole (~50%) and from electron transfer (~50%). The Solid lines are (bi)-exponential fits to the data.

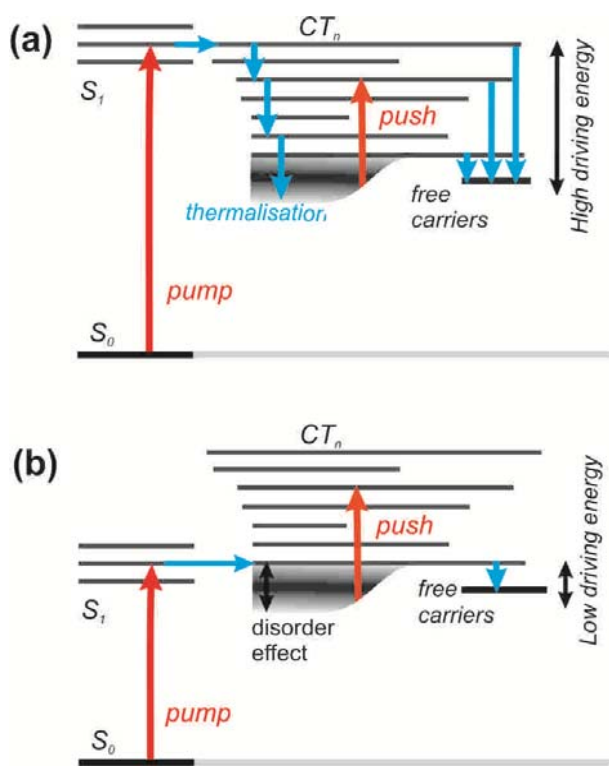

Figure 5. Potential energy diagrams describing charge separation in D-A molecular systems with high (a) and low (b) driving energy for charge separation. Red arrows are optical transitions used for spectroscopy. Blue arrows show relaxation-assisted sampling of different states after the initial excitation. Different width of depicted CT states reflects different delocalisation (and, therefore, tendency to dissociate) for those state. The static disorder effect on the position of the lowest CT level is shown by a diffused grey shape.

**TOC graphics**

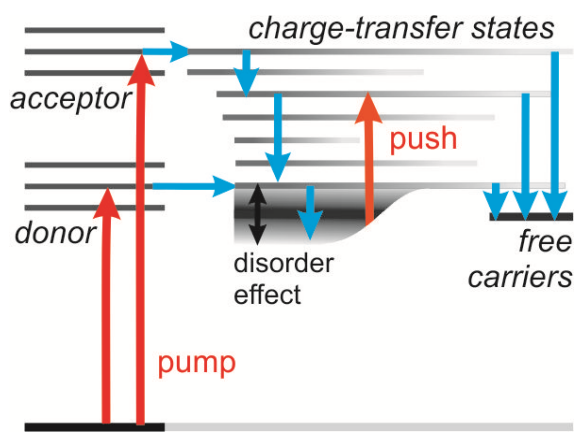

**Keywords:** ultrafast spectroscopy, conjugated polymers, charge separation, organic solar cells, IR spectroscopy, charge trapping